\documentclass{nature}
\usepackage{graphicx}
\title{An Anti-Glitch in a Magnetar}

\author{R. F. Archibald$^{1}$, V. M. Kaspi$^1$, C. -Y. Ng$^{1,2}$,   K. N. Gourgouliatos $^{1}$, D. Tsang$^{1}$,  P. Scholz$^{1}$,  A. P. Beardmore$^3$, N. Gehrels$^4$, \& J. A. Kennea$^5$}

\begin{document}

\maketitle
\begin{affiliations}
 \item Department of Physics, McGill University, Montreal QC, H3A 2T8, Canada
 \item Department of Physics, The University of Hong Kong, Pokfulam Road, Hong Kong 
 \item Department of Physics and Astronomy, University of Leicester, University Road, Leicester LE1 7RH, UK
 \item Astrophysics Science Division, NASA Goddard Space Flight Center, Greenbelt, MD 20771 USA
 \item Department of Astronomy and Astrophysics, 525 Davey Lab, Pennsylvania State University, University Park, PA 16802, USA
\end{affiliations}

\begin{abstract}
%\pagenumbering{gobble}
Magnetars are neutron stars showing dramatic X-ray and soft $\gamma$-ray outbursting 
behaviour that is thought to be powered by intense internal magnetic fields\cite{1996ThompsonMagnetar}.  Like
conventional young neutron stars in the form of radio pulsars, magnetars exhibit ``glitches'' during which
angular momentum is believed to be transferred between the solid outer crust and the superfluid
component of the inner crust \cite{1975AndersonGlitches, 1985Natur.316...27P, 1992NatureGLEOS}.  Hitherto, the several hundred observed glitches in radio pulsars\cite{2011EspinozaGlitches, 2013MNRAS.429..688Y} and magnetars\cite{2008Dib} have involved a sudden spin-up of the star, due presumably to the interior superfluid rotating
faster than the crust.  Here we report on X-ray timing observations of the magnetar
1E 2259+586 (ref. 8) %\cite{1981Fahlman})
 which we show exhibited a clear ``anti-glitch'' -- a sudden spin down.
We show that this event, like some previous magnetar spin-up glitches\cite{2003Kaspi2259}, was accompanied
by multiple X-ray radiative changes and a significant spin-down rate change.  
This event, if of origin internal to the star, is unpredicted in models of neutron star spin-down and is suggestive of
differential rotation in the neutron star, further supporting the need for a rethinking of glitch theory for all neutron stars.\cite{1998ApJ...492..267R,2012PhRvL.109x1103A}  
\end{abstract}

1E 2259+586 is a $\sim7$-s magnetar, with a spin-inferred surface dipolar magnetic field strength of 
$5.9 \times 10^{13}$ G. In 16 years of monitoring
with the {\it Rossi X-ray Timing Explorer (RXTE)}, 1E 2259+586 has shown a very stable 
spin-down rate and pulsed flux, with the exception of two spin-up glitches in
2002 (ref. 9) and 2007 (ref.12),  
%\cite{2003Kaspi2259})
%\cite{2012MNRAS.419.3109I}),
as well as a small timing event in 2009 (ref.12).%\cite{2012MNRAS.419.3109I}).
The 2002 glitch was also accompanied by
an increase in X-ray luminosity by a factor of 20 (ref. 9)% \cite{2003Kaspi2259})
 and X-ray bursts\cite{2004ApJ...607..959G}, neither of which was seen
in the 2007 glitch.

We began monitoring 1E 2259+586 with NASA's {\it Swift} X-ray Telescope (XRT)\cite{2005BurrowsSWIFT} in July 2011.  Observations have been made every 2-3 weeks, with typical exposure times of 4 ks. From each observation, we obtained a pulse time-of-arrival (TOA) by
folding the X-ray time series at the current pulse period and aligning this folded
light curve with a high signal-to-noise template. 

We fitted the pulse TOAs to a long-term timing model which keeps track
of every rotation of the neutron star. This
model predicts when the pulses should arrive at Earth, taking into account
the pulsar rotation as well as astrometric terms. We compared the observed TOAs with the
model predictions, and obtained best-fit parameters by $\chi^2$ minimization,
using the TEMPO2\cite{TEMPO2} software package. Until the observation on 14 April, 2012 (MJD
56,031.18)  these TOAs were well fitted using only a frequency and first frequency
derivative as shown in Fig.~1.

The subsequent data, however, clearly were not predicted by this simple model. TOAs starting on  28 April, 2012 (MJD 56,045.01) showed an apparently instantaneous change of the frequency -- which we dub an `anti-glitch.' On 21 April, 2012 (MJD 56,038), consistent with the epoch of this sudden spin down,  a 36-ms hard X-ray burst was detected by the {\it Fermi} Gamma-ray Burst Monitor (GBM)\cite{2012FoleyGCN}, with a fluence of $\sim6\times 10^{-8}$ erg/cm$^2$ in the $10 -1000$ keV range. No untriggered GBM bursts were seen within three days of the observed burst\cite{2012FoleyGCN}. As well, on 28 April, 2012 (MJD 56,045.01), coincident with the nearest  post-anti-glitch observation, 
we detected an increase in the 2 -10 keV flux by a factor of $2.00\pm0.09$ (see Fig. 1).
The 2-10 keV flux increase was also accompanied by a change in the hardness ratio, defined as
the ratio of the 4-10 keV to the 2-4 keV fluxes, from  $0.10\pm 0.02$ to $0.18 \pm 0.02$. This
flux increase subsequently decayed following a power-law model with $\alpha=-0.38\pm0.04$ (see Fig.~1).  The flux increase was accompanied by a moderate change in the pulse profile: the addition of a sinusoid centred between the usual two peaks in the pulse profile. This modified pulse profile relaxed back to the usual shape on a timescale similar to that of the flux. We verified that this profile change did not affect the TOAs determined near the anti-glitch epoch. 

This remarkable spin-down event was immediately followed by an extended period of enhanced spin-down rate. This anti-glitch and spin-down rate change can be well modelled by an instantaneous change in the frequency and frequency derivative, followed by a second sudden event.  We have found two possible timing models to describe the pulsar's behaviour, described in full in Table~1. In the first, there is an instantaneous change in frequency and frequency
derivative by $\Delta \nu = - 4.5(6)\times 10^{-8} $Hz  ($\Delta\nu/\nu =-3.1(4)\times 10^{-7}$) and $\Delta\dot{\nu} = -2.7(2)\times 10^{-14} $Hz/s  on $18$ April (MJD 56,035(2)).  This enhanced spin-down episode ended with a second glitch, this time a spin-up event, of amplitude  
$\Delta\nu =  3.6(7)\times 10^{-8} $Hz  ($\Delta\nu/\nu =2.6(5)\times 10^{-7}$)  and 
$\Delta\dot{\nu} = 2.6(2)\times 10^{-14} $Hz/s. 

In the second model, the spin evolution can be described by two anti-glitches, instead of an anti-glitch/glitch pair. In this model, a change of $\Delta \nu = - 9(1)\times 10^{-8} $Hz  ($\Delta\nu/\nu =-6.3(7)\times 10^{-7}$) and $\Delta\dot{\nu} = -1.3(4)\times 10^{-14} $Hz/s occurred on $21$ April (MJD 56,038(2)). This period ended with a second anti-glitch of amplitude  
$\Delta\nu =  -6.8(8)\times 10^{-8} $Hz  ($\Delta\nu/\nu =4.8(5)\times 10^{-7}$)  and 
$\Delta\dot{\nu} = 1.1(4)\times 10^{-14} $Hz/s. 

The full timing parameters for both possible models are presented in Table~1. Note that neither model is preferred on statistical grounds, however models involving a single initial anti-glitch and subsequent relaxation with no second impulsive event are ruled out to high confidence.  Also note that no significant radiative, or profile changes can be associated with either of the possible second impulsive events.

Note that in either model a sudden spin down at the epoch of the {\it Fermi} burst is unambiguously required to model the observed TOAs properly. While the amplitude of this anti-glitch in either model is not unusual, the fact that it is a sudden spin down is remarkable. The net effect of this active period are changes to the spin frequency and its first derivative $\Delta \nu = - 2.06(8)\times 10^{-7} $Hz  ($\Delta\nu/\nu =-1.44(6)\times 10^{-6}$) and a $\Delta\dot{\nu} = -1.3(4)\times 10^{-15} $Hz/s.

Sudden spin-down glitches have heretofore not been observationally demonstrated,
though some magnetar events have been suggestive.   A spin-down in magnetar
SGR 1900+14 (ref.17) %\cite{1999Woods1900Spinup})
 occurred during an 80-day gap in the source monitoring, but could
have been a gradual slow down, as was also possible for the 2009 timing event in
1E 2259+586 (ref.12). %\cite{2012MNRAS.419.3109I}).
Net spin downs have been seen in magnetar 4U 0142+61 (ref.18)  %\cite{2011Gavrill0142})
 and
in the high magnetic field rotation-powered pulsar PSR J1846$-$0258 (ref.19)% \cite{2010Livingstone1846})
 but
were due to spin-up glitch over-recoveries on time scales of 17 and 127 days, respectively.
If the 1E 2259+586 event were due to a spin-up glitch and subsequent over-recovery,
we place a $3\sigma$ upper limit on the recovery decay time of 3.9 days for a spin-up
of size $\Delta \nu/\nu = 1 \times 10^{-6}$.  Even for an infinitesimally small spin-up
glitch, the decay time was less than 6.6 days, far shorter than any previously observed
magnetar recovery time scales.

Following the detection of the anti-glitch, we looked for evidence of particle outflow, proposed as a possible mechanism for the apparent spin down in SGR 1900+14 (ref. 20). %\cite{2000ThompsonSpinDown}).
 We carried out radio imaging on  21 August,  2012 using the Expanded Very Large Array in the B-array configuration with a 240-minute integration time. This yielded images with
effective angular resolution 1.2$''$. We performed standard flagging, calibration, and imaging using
the Common Astronomy Software Applications (CASA) package\cite{2011ascl.soft07013I}.
No source was found at the position of 1E 2259+586, and we place a
$3\sigma$ flux density limit of 7.2\,$\mu$Jy at 7\,GHz for a point source. This is significantly lower than the previous upper limit of $50$  $\mu$Jy at 1.4 GHz\cite{2003Kaspi2259}. If a putative outflow were expanding at $0.7c$ as was the case for a radio outflow from SGR 1806$-$20    (ref.22)% \cite{2006ApJ...638..391G}) 
at the time of its outburst, we would expect a nebular radius of $4''$.  For this radius, we
obtain a $3\sigma$ flux density limit of 0.46\,mJy.
Note that the limit is more stringent if the size is smaller, and reduces to
7.2\,$\mu$Jy  if unresolved.

In X-rays, we also detected no evidence for such outflow in a 10-ks {\it Chandra} HRC-I  image taken on  21 August,  2012.
Using simulations, we  place an upper limit 
on  X-ray flux from a putative outflow at $2\%$ of the total 1-10 keV X-ray emission of the
magnetar, for a $4''$ circular nebula with a Crab-like spectrum.

There are two main possibilities for the origin of the anti-glitch: either an internal or external mechanism. 

An impulse-like angular momentum transfer between regions of more slowly spinning superfluid and the crust could be the source of the anti-glitch.\cite{2000ThompsonSpinDown} A slower angular momentum transfer to such a region or the decoupling of a significant amount of the moment of inertia of the star could account for the enhanced spin-down rate. The second event, either glitch or anti-glitch, can similarly be modelled by angular momentum transfers from differentially rotating regions of the neutron star superfluid. The radiatively quiet nature of the second event does not pose a problem for the internal model as many glitches are radiatively silent\cite{2008Dib}. The behaviour indicated by an impulsive anti-glitch offers new evidence for possible significant internal structural changes and differential rotation in magnetars at glitch epochs.

An external model such as an outflow along the open field lines of the magnetosphere\cite{1998PhRvD..57.3219T, 1999ApJ...525L.125H, 2000ThompsonSpinDown}, or a sudden twisting of the field lines\cite{2009ApJ...703.1044B} can be the cause of the anomalous spin-down behaviour. However, in a wind model, the second timing event should also be accompanied by a radiative change, as the first one. If this behaviour was caused by twisting magnetic field lines, it should be followed by a gradual untwisting and a similar behaviour reflected in $\dot{\nu}$.\cite{2012ParfreyMagn} (see Supplemental Material)

Overall, this magnetar anti-glitch, X-ray outburst, and subsequent evolution lend additional support to the need for a rethinking of glitch theory for all neutron stars.\cite{1998ApJ...492..267R, 2012PhRvL.109x1103A}

\begin{table}

\caption{Timing parameters for 1E 2259+586. Numbers in parentheses are TEMPO-reported 1 $\sigma$ uncertainties.}
\begin{tabular}{ c c }
\hline
\hline
Parameter & Value \\
\hline
Observation Dates & 23 July 2011 - 1 January 2013 \\
Dates (MJD) & $55,765.829 - 56,293.332$\\
Epoch (MJD) & $55,380.000$\\
Number of TOAs & $51$\\
$\nu$  (s$^{-1}$)  & $0.143,285,110(4)$\\
$\dot{\nu}$ (s$^{-2}$) & $-9.80(9) \times 10^{-15}$  \\
\multicolumn{2}{c}{Model 1} \\
Glitch Epoch 1 (MJD)  & $56,035(2)$\\ 
$\Delta \nu_1$ ($s^{-1}$)   & $-4.5(6)  \times 10^{-8}$\\
$\Delta \dot{\nu_1}$  ($s^{-2}$)   & $-2.7(2)  \times 10^{-14}$\\
Glitch Epoch 2 (MJD)  & $56,125(2)$\\ 
$\Delta \nu_2$  (s$^{-1}$)  & $3.6(7)  \times 10^{-8}$\\
$\Delta \dot{\nu_2}$ (s$^{-2}$)  & $2.6(2)  \times 10^{-14}$\\
RMS residuals (ms)  & $56.3$\\
$\chi^{2} / \nu$  & $45.4/44$\\
\multicolumn{2}{c}{Model 2} \\
Glitch Epoch 1 (MJD)  & $56,039(2)$\\ 
$\Delta \nu_1$ ($s^{-1}$)   & $-9(1)  \times 10^{-8}$\\
$\Delta \dot{\nu_1}$  ($s^{-2}$)   & $-1.3(4)  \times 10^{-14}$\\
Glitch Epoch 2 (MJD)  & $56,090(3)$\\ 
$\Delta \nu_2$  (s$^{-1}$)  & $-6.8(8)  \times 10^{-8}$\\
$\Delta \dot{\nu_2}$ (s$^{-2}$)  & $1.1(4)  \times 10^{-14}$\\
RMS residuals (ms)  & $51.5$\\
$\chi^{2} / \nu$  & $38.1/44$\\
\end{tabular}

\end{table}

\begin{figure}
\caption{Timing and flux properties of 1E 2259+586 around the 2012 event. Panel {\textbf a} shows 1E 2259+586's spin frequency as a function of time, determined by short-term fitting of typically 5 TOAs. The grey horizontal error-bars indicate the ranges of dates used to fit the frequency, and the vertical error bars (generally smaller than the points) are standard $1 \sigma$ uncertainties. The red and blue solid lines in panel {\textbf a} represent the fits to the pulse TOAs, as displayed in Table 1, with red representing model 1, and blue model 2. Panel {\textbf b} shows the timing residuals of 1E 2259+586 after fitting only for the pre-anti-glitch timing solution. The inset shows the same timing residuals, zooming in on the anti-glitch epoch.   Panel  {\textbf c} shows the absorbed 2-10 keV X-ray flux. The error bars indicate the $1 \sigma$ uncertainties, and the green line is the best-fit power-law decay curve with an index of $-0.38\pm0.04$.The dashed vertical lines running through both panels indicate the glitch epochs, the black being the anti-glitch, and blue and red the second event in the models shown in Table~1. The timing residuals for these fits can be seen in the supplementary material.}
\includegraphics[width=\textwidth]{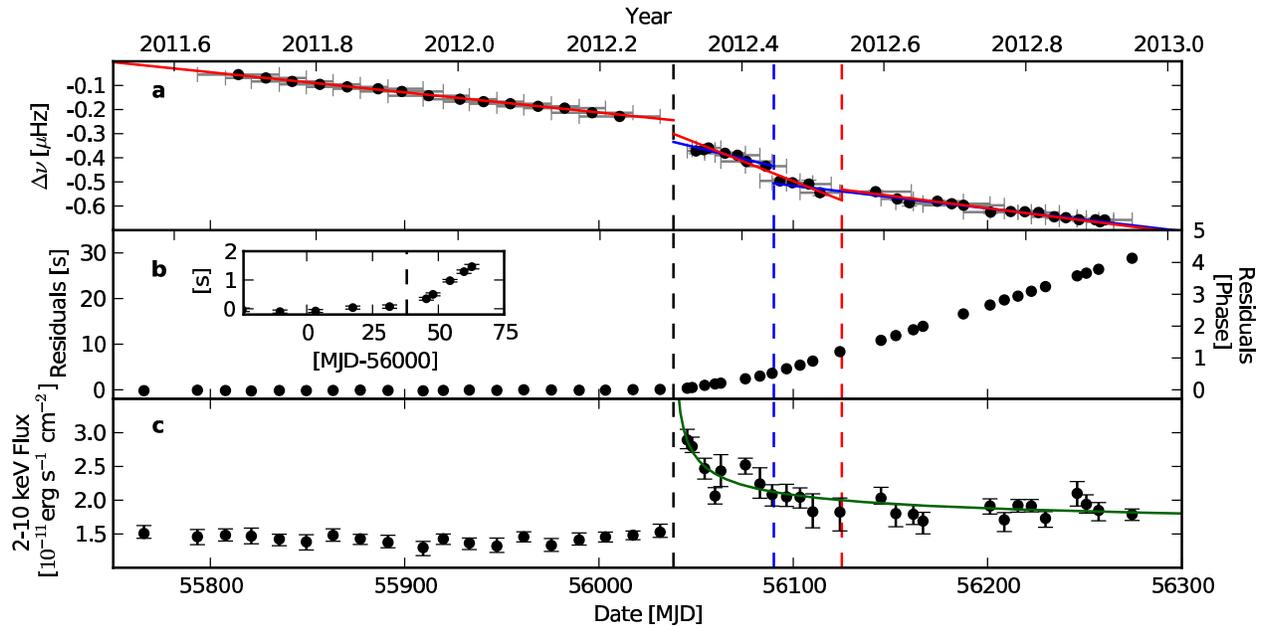}
\end{figure}

\clearpage
%\pagenumbering{arabic}
%Supplementary information
%\doublespace
\section{Observations}
The {\it Swift} X-Ray Telescope (XRT) is a Wolter-I
telescope with an {\it XMM-Newton} EPIC-MOS CCD22 detector, sensitive in the 0.5 -10 keV range. The XRT was operated in Windowed-Timing (WT) mode for all observations, which gives  1.76-ms time
resolution. {\it Swift} observations of 1E 2259+586 had typical exposure times of 4 ks.

The X-ray flux was measured by processing level 1 data products which were obtained from the HEASARC {\it Swift} archive, reduced using the {\it xrtpipeline} standard reduction script, and barycentred to the position of 1E 2259+586 ($RA= 23^h01^m07.900^s, DEC=58^\circ  52' 46.00''$), using HEASOFT $v6.12$ and  the {\it Swift} 20120209 CALDB. A 40-pixel long region centred on the source was extracted, as well as a background region of the same size located away from the source.  
To investigate the flux and spectral behaviour of 1E 2259+586, spectra were extracted from the selected regions using {\it xselect}, and fit to an $N_H$-absorbed blackbody and power-law model using { \it XSPEC} package version 12.7.0q\cite{1996ASPC..101...17A}. $N_H$ was fixed at the value of $0.97\pm 0.03 \times 10^{22}$ cm$^{-2}$, determined by co-fitting all the pre-glitch spectra. The spectra were grouped with a minimum of 20 counts per energy bin.  Ancillary response files were created using the FTOOLS {\it xrtmkarf} and the standard spectral redistribution matrices. 

\section{Discussion}

The physical cause of the glitching behaviour of the magnetar and the enhanced spin down can be due to either internal or external mechanisms.

In regular pulsars, we expect the crust to spin slower than the uniformly rotating superfluid. However, in a magnetar, there is significant internal free energy generated by the magnetic-field decay\cite{1996ThompsonMagnetar}, which could potentially drive differential rotation. Such differential rotation could allow for regions of superfluid to be spinning slower than the crust. An impulse-like angular momentum transfer between such regions and the crust could be the source of the anti-glitch, while a slower angular momentum transfer to such a region could account for the enhanced by a factor of either $\sim 4$ or $\sim 2$ 
 spin-down rate. Another possible cause of the enhancement in the spin-down rate is by decoupling $\sim 3/4$ or $\sim 1/2$, respectively, of the moment of inertia on which the torque is acting\cite{2003Kaspi2259}. In a normal pulsar glitch, $\sim 1\%$ of the moment of inertia is required to explain the observed $\Delta\dot{{\nu}}/{\nu}$ ̇ of about 0.01. Recent studies\cite{2013PhRvL.110a1101C, 2012PhRvL.109x1103A} suggest that even in normal pulsars, the crustal superfluid does not provide sufficient angular momentum for glitches, and a larger reservoir is needed. The extreme fields in magnetars could mediate the coupling of the crust and superfluid in the outer core. Events that release energy to the outside in the form of bursts should also alter the internal dynamics, leading to changes in the coupling and eventually to the moment of inertia. The exchange of energy between the core and the crust can heat the latter and enhance the X-ray luminosity\cite{2012ApJ...750L...6P}.

Another internal mechanism that has been proposed in the context of SGR 1900+14\cite{2000ThompsonSpinDown} is the twisting of a patch of crust by a magnetospheric event which could cause a change in the angular momentum of the superfluid by the net motion of pinned vortices. Such a mechanism would typically cause a net spin up of the crust for a uniformly rotating superfluid. In this model, a slow plastic twist resulting in a sudden unpinning event could cause an anti-glitch. However, such a slow plastic deformation would be accompanied by a small relative decrease in the spin-down rate prior to the anti-glitch, which was not observed.

An external mechanism which varies the torque could also have caused the observed spin-down behaviour. An outflow along the open field lines of the magnetosphere could provide an additional torque which would increase the spin down by a factor of $\sim 2 - 4$. This would require a wind of luminosity $\sim1.5 \times 10^{33} $ erg s$^{-1}$ to act for a week, to explain the initial anti-glitch and to be followed by a $\sim 1 \times 10^{32}$ erg s$^{-1}$ wind to cause the enhanced spin-down. In the two anti-glitch model, it would conclude with a rejuvenated flux to explain the increased torque at the time of the second anti-glitch. While the levels of X-ray luminosity due to the enhanced spin down would be undetectable in our monitoring, that from the short-term strong winds needed to explain the anti-glitches would have been. While we cannot exclude this model, it has low predictive power. Similarly, a sudden twisting of the field lines, through internal magnetic evolution, or external field activity, during the initial event can lead to larger torques. The twist needed to achieve this torque is $\sim 3$ rad if the displacement is confined in the polar cap or in a ring in that region,\cite{2009ApJ...703.1044B} and $\sim 1$ rad for a global twist\cite{2002ApJ...574..332T}. However, if the currents supporting the twisted field dissipate smoothly following the initial twist, then $\dot{\nu}$ should also follow this trend. In this model the second anti-glitch requires a similar process. While a clear X-ray outburst and profile change were detected coincident with the first anti-glitch, there was no significant increase in X-ray luminosity, nor significant pulse profile changes coinciding with the second anti-glitch, even though the two events would have been similar in magnitude.

Determining whether the observed anti-glitch had origin internal or external to the star clearly has potential importance for our understanding of neutron-star structure.  This could be accomplished in principle by, for example, better constraining the time scale on which the anti-glitch occurs, since an internal angular momentum transfer is likely to yield a near-instantaneous event, where as magnetospheric twists should have a longer evolution time scale.  A sensitive all-sky X-ray monitor would be useful in this regard.

%\singlespace
\newpage

\begin{figure}
\caption{ 
%{\bf Supplementary Figure 1:} 
Timing residuals for 1E 2259+586. Panels {\textbf a},  {\textbf b}, and  {\textbf c} show timing residuals, the difference between the predicted and measured TOAs for the timing models shown in Table~1. Panel {\textbf a} takes into account only the long-term pre-anti-glitch timing model, with the inset showing the phase jump and significant slope change that indicates the anti-glitch. Panel {\textbf b} shows these residuals after fitting Model 1, the anti-glitch and the glitch, and Panel {\textbf c} for Model 2, the two anti-glitch model. The dashed vertical lines indicate glitch epochs -- the black line the common anti-glitch epoch, the red line  the glitch epoch in model 1, and the blue the second anti-glitch epoch in model 2. } 

\includegraphics[width=\textwidth]{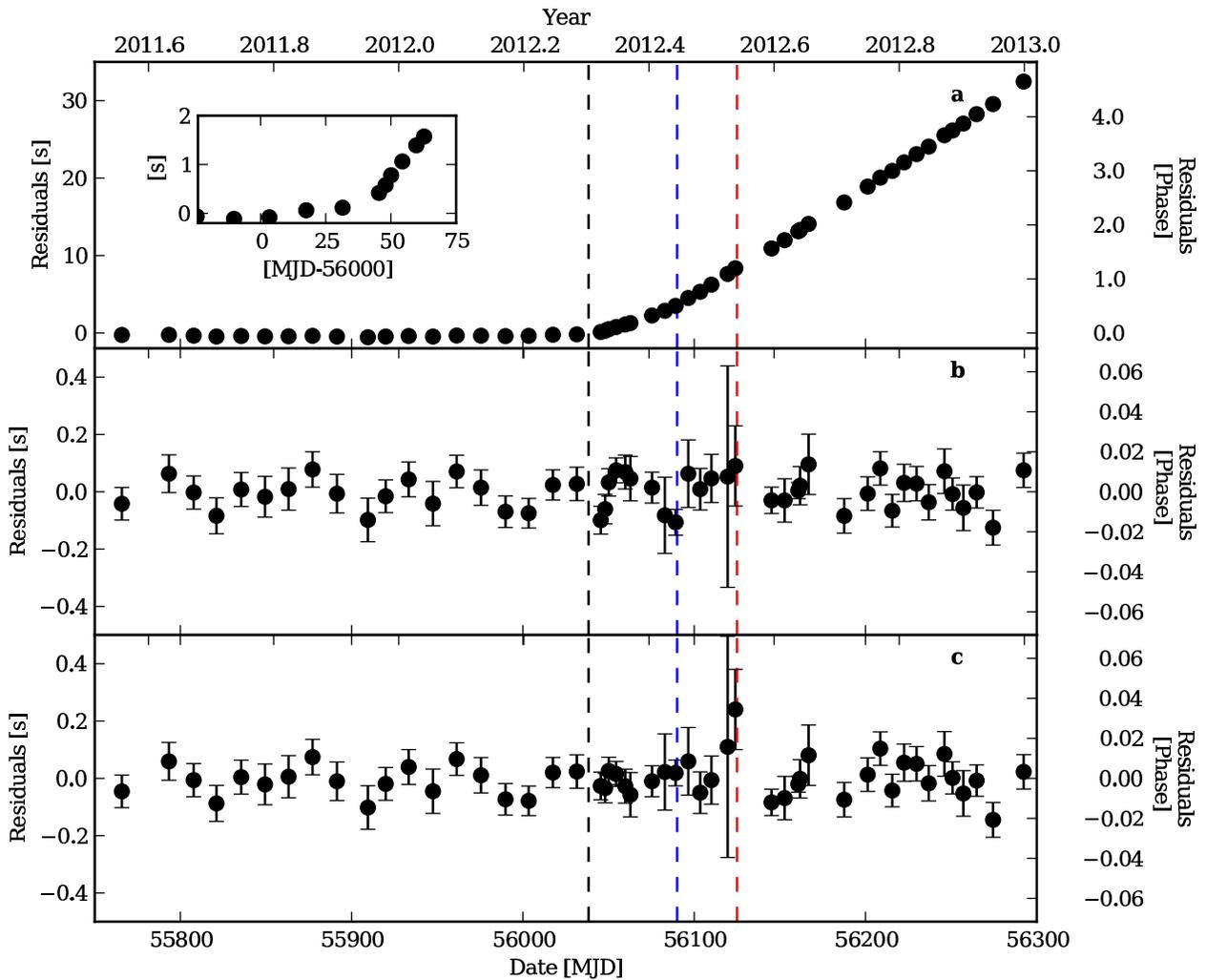}
\label{Flux}
\end{figure}

\newpage
\bibliography{Merged}{}

{\bf Acknowledgements} V.M.K. acknowledges support from the Natural Sciences and Engineering Research Council of Canada Discovery Grant and John C. Polanyi Award, from the Canadian Institute for Advanced Research, from Fonds de Recherche Nature et Technologies Qu\'ebec, from the Canada Research Chairs Program, and from the Lorne Trottier Chair in Astrophysics and Cosmology. D.T. was supported by the Lorne Trottier Chair in Astrophysics and Cosmology and the Canadian Institute for Advanced Research. K.N.G. was supported by the Centre de Recherche en Astrophysique du Qu\'ebec. We thank Heidi Medlin and Joseph Gelfand for help with the EVLA observation. We thank D. Eichler, B. Link, M. Lyutikov and C. Thompson for useful discussions. We acknowledge the use of public data from the Swift data archive 

{\bf Contributions} R.F.A. performed the data analysis and wrote portions of the analysis software.  V.M.K. designed the study, was the project leader for the {\it Swift} data, proposed for the {\it Chandra} data, assisted with the interpretation of the data analysis, as well as with the theoretical implications.  C.Y.N. proposed for the VLA data and reduced them and the {\it Chandra} data.  K.N.G. and D.T. assisted with the theoretical implications.  P.S. wrote significant portions of the {\it Swift} analysis software. A.P.B., N.G. and J.K. assisted with {\it Swift} observations and data analysis.  R.F.A. wrote the paper with guidance from V.M.K. and with significant input from all co-authors. 

{\bf Competing Interests} The authors declare that they have no competing financial interests. 

{\bf Correspondence} Correspondence and requests for materials should be addressed to (V.M.K., email:  vkaspi@physics.mcgill.ca).

\end{document}